\documentclass{article}

\pdfoutput=1
\usepackage{Others/PRIMEarxiv}

\usepackage[utf8]{inputenc} 
\usepackage[T1]{fontenc}    
\usepackage{url}            
\usepackage{booktabs}       
\usepackage{amsfonts}       
\usepackage{amsmath}
\usepackage{nicefrac}       
\usepackage{microtype}      
\usepackage{lipsum}
\usepackage{fancyhdr}       
\usepackage{graphicx}       
\graphicspath{{Figures/}}     
\DeclareUnicodeCharacter{0301}{ }

\usepackage{biblatex}
\usepackage[pdftex]{hyperref}

\title{$\omega$-FWI: Robust full-waveform inversion with Fourier-based metric}

\author{
  Muhammad Izzatullah\thanks{\textbf{Corresponding author:} \texttt{muhammad.izzatullah@kaust.edu.sa}} , Tariq Alkhalifah \\
  Seismic Wave Analysis Group (SWAG) \\
  King Abdulllah University of Science and Technology (KAUST) \\
}

\begin{document}
\maketitle

\begin{abstract}
Full-waveform inversion is a cutting-edge methodology for recovering high-resolution subsurface models. However, one of the main conventional full-waveform optimization problems challenges is cycle-skipping, usually leading us to an inaccurate local minimum model. A highly investigated track to alleviate this challenge involves designing a more global measure of misfit between the observed and modelled data beyond the sample-to-sample comparison. However, most of these approaches admit relatively smooth inversion results. Here, we introduce a novel misfit function based on the \emph{Fourier-based metric}. This metric has been successfully applied in molecular physics for solving the Boltzmann equation, and we adapt it to full-waveform inversion. This misfit function exploits the power spectrum information between the modelled and observed data to provide low-wavenumber velocity model updates early, and more high resolution updates as we approach the solution. Thus, it also can be reformulated as a weighted $\ell_{2}$-norm in a quadratic case, which can be seen as a simple extension for conventional full-waveform inversion. Thus, despite its robustness to cycle skipping, it is capable of delivering high-resolution models synonymous to conventional FWI. Considering its frequency domain utilization, we refer to this inversion method as \emph{$\mathbf{\omega}$-\textbf{FWI}}. Through the synthetic Marmousi model example, this method successfully recovers an accurate velocity model, starting from a linearly increasing model even for the case of noisy observed data and the lack of low frequencies below 3 Hz and 5Hz, in which the conventional $\ell_{2}$-norm full-waveform inversion suffers from cycle skipping.
\end{abstract}

\section{Introduction}
Full-waveform inversion (FWI) addresses the geophysical inverse problem of recovering a high-resolution subsurface model from observed data through a nonlinear optimization procedure. This optimization procedure iteratively updates the subsurface model by minimizing a misfit function that characterizes the mismatch between the modelled and observed data. The foundation of FWI was introduced by \cite{Lailly1983} and \cite{Tarantola1984} based on the $\ell_{2}$ norm. However, this formulation leads FWI into a local minimum when the initial model is not located in the convergence region of the global minimum. Therefore, considerable effort have been devoted to building a promising misfit function with better convexity property. A well-designed misfit function would relax the requirement for a kinematically accurate initial model or the necessity of low and ultra-low frequencies in the observed data to achieve a successful inversion and resolve the so-called cycle skipping issue.

Recently, more advanced misfit functions have been proposed, such as the optimal transport \cite{Yang2018, Metivier2018, Yang2018b, Chen2018, Chen2018b}, differentiable dynamic time warping \cite{Chen2021, Chen2021b}, double-difference \cite{Chen2022}, matching filter \cite{Warner2016, Sun2019, Sun2019b} and deep-learning \cite{Yang2021} based misfit functions. Instead of characterizing the mismatch locally (sample-by-sample comparison), those newly proposed methods focus on characterizing the mismatch in their global features, involving the full trace or even the full gather. By characterizing the mismatch globally, the resulting misfit function shows more convex behaviour, as it reasonably mitigates the nonlinearity in FWI. However, these global data misfit measures often admit lower resolution velocity models, and thus, they are often followed up by an application of an $\ell_{2}$-norm FWI. For recent comparison studies performed on several recent proposed misfit functions, we refer you to \cite{Pladys2021}.

In this work, we introduce a novel misfit function that utilizes a \emph{Fourier-based metric}. This metric has been successfully applied in molecular physics for solving the Boltzmann equation \cite{Gabetta1995}. In the FWI application, this  misfit function exploits the power spectrum information between the modelled and observed data to provide the best velocity model updates. We also demonstrate that this misfit function can be reformulated as a weighted $\ell_{2}$-norm in a quadratic case, which can be seen as a simple extension of the conventional full-waveform inversion and thus, maintains its high-resolution properties. We refer to this inversion methodology as \emph{$\mathbf{\omega}$-\textbf{FWI}}.

The outline of this work is as follows. First, we discuss the theoretical foundation of the proposed misfit function. Next, we demonstrate the features of this methodology through a convexity analysis of the misfit function and numerical examples using the infamous Marmousi model on time-domain FWI. Finally, we share our conclusions.
\section{Theoretical Framework}
\subsection{The Fourier-based metric}

The Fourier-based metric, also known as the \emph{Tanaka metric}, was first introduced in \cite{Gabetta1995} based on the work of \cite{Tanaka1973, Tanaka1978} to study the large-time asymptotics of Boltzmann equation for Maxwellian molecules. In recent years, several investigations were performed to relate the optimal transport theory to Fourier-based metric \cite{Carrillo2007cpm, Auricchio2020, Steinerberger2021}. In this work, we generalize the Fourier-based metric and present its connection with the weighted $\ell_{2}$-norm for FWI applications. We refer to the application of this metric to FWI as \emph{$\mathbf{\omega}$-\textbf{FWI}}.

For a given waveform signal $\mathbf{d}(t) \in \mathrm{R}^{N_{t}}$ where $N_{t}$ is the number of time samples, we define the Fourier transform of $\mathbf{d}(t)$ as

\begin{equation*}
    \Hat{\mathbf{d}}(\omega) = \int_{\mathrm{R}^{N_{t}}} \mathbf{d}(t) \exp{(-\mathrm{i} \omega t)} dt
\end{equation*}
where $\omega$ and $t$ represent the angular frequency and time, respectively. By taking a Fourier transform on two given waveform signals $\mathbf{d}_{m}(t), \mathbf{d}_{o}(t) \in \mathrm{R}^{N_{t}}$, the \emph{Fourier-based metric} can be defined as
\begin{equation}\label{eq:1}
    \mathcal{L}_{p, \alpha}(\mathbf{d}_{m}, \mathbf{d}_{o}) = \Bigg( \int |\omega|^{\alpha} |\Hat{\mathbf{d}_{m}}(\omega) - \Hat{\mathbf{d}_{o}}(\omega)|^{p} d\omega \Bigg)^{1/p}
\end{equation}
where $p, \alpha \in \mathrm{R}$, and $\Hat{\mathbf{d}_{m}}$ and $\Hat{\mathbf{d}_{o}}$ are the Fourier transforms of the modelled $\mathbf{d}_{m}$ and observed $\mathbf{d}_{o}$ waveforms, respectively. For FWI applications, this metric can be seen as a trace-by-trace comparison. However, the extension to a higher dimension (e.g., gather by gather) can be accomplished with $N$-dimensional Fourier Transforms \cite{Tolimieri2012}. 

Intuitively, this Fourier-based metric allows us to control how much a given order of derivative (or integration in the case of $\alpha < 0$) in the waveform can contribute to the mismatch between the modelled and observed data in the Fourier domain. This approach also can be seen as a data preconditioner by simply allowing the low frequencies in the data to be enhanced in an edge-preserving manner before the $\ell_{2}$-norm computation takes place. 

Now, we demonstrate the connection between this metric and the weighted $\ell_{2}$-norm, and specifically conventional FWI is a specific case of the new metric measure. For $p=2$, we can express the Fourier-based metric in the equation \ref{eq:1} in a weighted $\ell_{2}$-norm formulation as 

\begin{equation}\label{eq:2}
\begin{split}
    \mathcal{L}_{2, \alpha}(\mathbf{d}_{m}, \mathbf{d}_{o}) &= (\Hat{\mathbf{d}_{m}} - \Hat{\mathbf{d}_{o}})^{T} \mathrm{W}_{\alpha} (\Hat{\mathbf{d}_{m}} - \Hat{\mathbf{d}_{o}})\\
    &= (\mathbf{d}_{m} - \mathbf{d}_{o})^{T} \mathrm{\Omega}^{T}\mathrm{W}_{\alpha}\mathrm{\Omega} (\mathbf{d}_{m} - \mathbf{d}_{o})\\
\end{split}
\end{equation}
where $\mathrm{\Omega}$ is the \emph{Discrete Fourier Transform (DFT)} matrix, and $\mathrm{W}_{\alpha} = \text{diag}(|\mathbf{\omega}|^{\alpha})$, where $\mathbf{\omega}$ is a vector of waveform's sample frequencies. Thus, the matrices $\mathrm{\Omega}$ and $\mathrm{W}_{\alpha}$ act as weighting functions on the $\ell_{2}$-norm allowing for scaling the power spectrum in the Fourier domain. Based on the formulation in equation \ref{eq:2}, the gradient and Hessian matrix of the Fourier-based metric for $p=2$ with respect to $\mathbf{d}_{m}$ can be expressed explicitly as:

\begin{equation}\label{eq:3}
    \nabla_{\mathbf{d}_{m}}\mathcal{L}_{2, \alpha} = 2 \mathrm{\Omega}^{T}\mathrm{W}_{\alpha}\mathrm{\Omega} (\mathbf{d}_{m} - \mathbf{d}_{o}),
\end{equation}
and

\begin{equation}\label{eq:4}
    \nabla^{2}_{\mathbf{d}_{m}}\mathcal{L}_{2, \alpha} = 2 \mathrm{\Omega}^{T}\mathrm{W}_{\alpha}\mathrm{\Omega}.
\end{equation}
With these explicit formulations of misfit gradient and Hessian, we can easily implement them in a computational graph fashion to leverage the advantages of automatic differentiation as presented in a deep-learning framework such as PyTorch \cite{Paszke2019}.\\


\subsection{$\mathbf{\omega}$-FWI problem formulation}

Full-waveform inversion aims to estimate the subsurface model $\mathbf{m}$ by minimizing the misfit between the modelled data $\mathbf{d}_{m}(\mathbf{m})$ and observed data $\mathbf{d}_{o}$. Here, we focus on the Fourier-based metric with $p=2$ as presented in the equation \ref{eq:2}, and we define our misfit function as

\begin{equation}\label{eq:5}
    \min_{\mathbf{m}} \mathrm{J}(\mathbf{m}) = \frac{1}{2} (\Hat{\mathbf{d}_{m}} - \Hat{\mathbf{d}_{o}})^{T} \mathrm{W}_{\alpha} (\Hat{\mathbf{d}_{m}} - \Hat{\mathbf{d}_{o}}),
\end{equation}
where the modelled data $\mathbf{d}_{m}$ is obtained through simulating the propagation of seismic waves by solving an acoustic wave equation and extracting the wavefield at the receivers level.

We update the subsurface model $\mathbf{m}$ iteratively using a gradient-based optimization method to minimize the misfit function for a specific choice of $\alpha$. The gradient can be computed using the adjoint-state method \cite{Plessix2006} by cross-correlating the forward-propagated wavefield (state variable) with the back-propagated wavefield (adjoint-state variable) from the adjoint source in the equation \ref{eq:3}.
\section{Numerical Examples}
\subsection{Misfit function convexity analysis}

\begin{figure}
    \centering
    \includegraphics[width=0.6\textwidth]{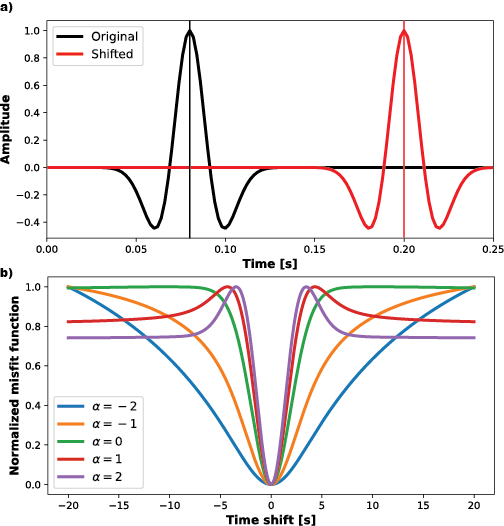}
    \caption{Misfit function convexity analysis: \textbf{(a)} The original and shifted signals, \textbf{(b)} the misfit function value as a function of time-shift for different $\alpha$.}
    \label{Convexity}
\end{figure}

We analyse the convexity properties of the proposed misfit function as $\alpha$ varies. In Figure \ref{Convexity}(a), we use two Ricker wavelets with a dominant frequency of $20$ Hz to represent the original and time-shifted versions. Here, the time shift ranges from $-20.0$ to $20.0$ s. We compare the results obtained with different values of $\alpha$ in Figure \ref{Convexity}(b). 

In this setup, we can observe that as $\alpha$ decreases to a negative value, we obtain good convexity, indicating that negative values of $\alpha$ can better mitigate the nonlinearity of FWI, especially with respect to cycle skipping. However, we discovered, yet not presented in this work, that very low $\alpha$ (e.g., $\alpha < -4$) tends to admit strong convexity, resulting in very smooth gradients. This phenomenon is equivalent to applying Gaussian smoothing on the $\ell_{2}$-norm misfit gradient update, and in the limit of strong smoothing, all high-wavenumber features are lost \cite{Zuberi2018}. Thus, we set $\alpha = -2$ and demonstrate its robustness in the following example.

\subsection*{Marmousi model}

\begin{figure}
    \centering
    \includegraphics[width=1\textwidth]{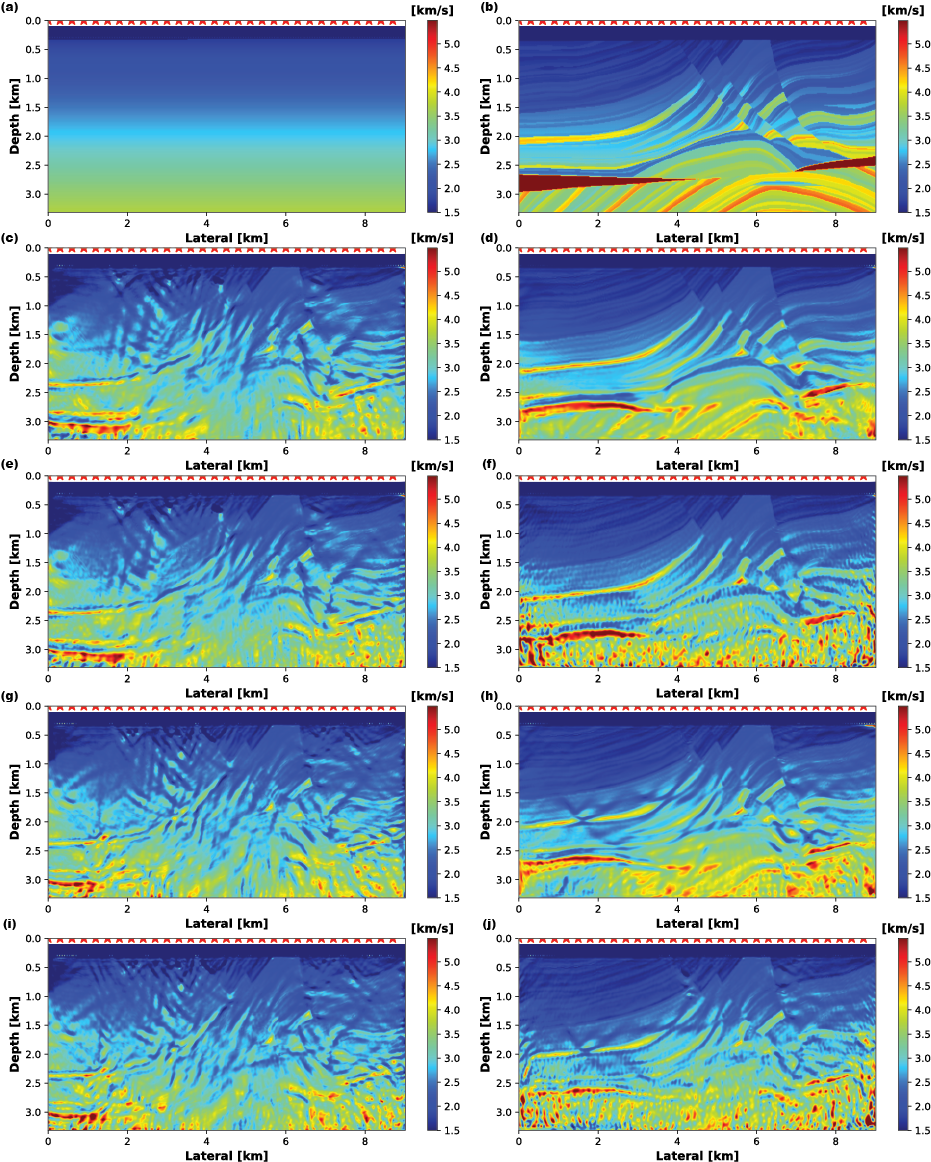}
    \caption{\textbf{a)} The initial model, \textbf{b)} the true model. The first column are results from the $\ell_{2}$-norm and the second column from the proposed method $\mathbf{\omega}$-FWI. The inverted model from clean observed data in the absence of frequencies below 3Hz and 5Hz, respectively: \textbf{(c)}\textbf{(g)} $\ell_{2}$-norm, and \textbf{(d)}\textbf{(h)} $\mathbf{\omega}$-FWI. The inverted model from noisy observed data in the absence of frequencies below 3Hz and 5Hz, respectively: \textbf{(e)}\textbf{(i)} $\ell_{2}$-norm, and \textbf{(f)}\textbf{(j)} $\mathbf{\omega}$-FWI. The red stars and white triangles represent the sources and receivers, respectively.}
    \label{Inversion}
\end{figure}

This classic (Marmousi model) example allows us to investigate the robustness of the proposed inversion method in the time domain through what we have seen of the inversion of this model using other methods. The initial velocity is a $v(\mathbf{z})$ model as illustrated in Figure \ref{Inversion}(a). The observed data are modelled using 30 shots with a source interval of $300$ m and $300$ receivers with an interval of $30$ m. The source is a Ricker wavelet with a $10$ Hz dominant frequency. The maximum recording time is $4$ s. 

\begin{figure}
    \centering
    \includegraphics[width=1\textwidth]{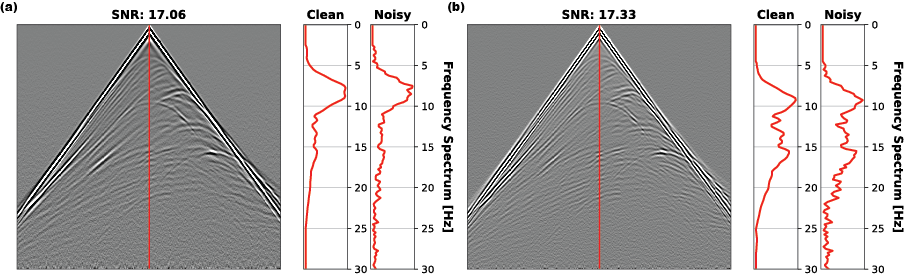}
    \caption{Noisy observed data and its clean and noisy power spectrum at seismic trace $150$. We mute data below \textbf{a)} 3Hz and \textbf{b)} 5Hz, respectively.}
    \label{Data}
\end{figure}

We compare the proposed approach directly with the conventional $\ell_{2}$-norm FWI for 2 different scenarios. We mute data below $3$ and $5$ Hz for each scenario. In the first scenario, we compare the robustness of both misfit functions in mitigating the cycle skipping issue by inverting clean observed data starting from a linearly increasing model in the absence of low-frequency information. Next, to replicate a realistic scenario in the inversion, we add coloured noise to the observed data within the same frequency spectrum. The resulting noisy observed data has a signal-to-noise ratio (SNR) of $17.06$ and $17.33$, as illustrated in Figure \ref{Data}. 
Here, we perform the inversion with an Adam optimizer \cite{Kingma2014}. The inversion is performed using Deepwave\footnotemark{}, the wave propagation modules for PyTorch \cite{deepwave} \footnotetext{Deepwave: \url{https://github.com/ar4/deepwave}}.

For the first scenario, the inverted results for the $\ell_{2}$-norm are illustrated in Figures \ref{Inversion}(c)(g), and for the proposed method are illustrated in Figures \ref{Inversion}(d)(h). Both results correspond to the absence of low frequencies below $3$ and $5$ Hz, respectively. Compared to the true model, the proposed method successfully recovered the global features and fine-scale parts of the Marmousi model, even though we are starting far from the true model. 
On the $\ell_{2}$-norm side, it generally recovered the global features of the Marmousi model. However, it failed to converge to a well-defined model due to cycle-skipping. On the other hand, the proposed method can mitigate the cycle skipping issue. To explain this result, intuitively, the proposed method is minimizing the energy difference between two waveforms instead of minimizing the amplitudes. Also, with the scaling properties of $\mathrm{W}_{\alpha}$ in the equation \ref{eq:5}, it boosted the low-frequencies in the waveform to provide better model updates; this is only correct for cases where $\alpha < 0$. As we get closer to the true solution (global minimum) the updates admit higher resolution consistent with Figure \ref{Convexity}(b). In other words, as the low-frequency components of the data match, the role of the frequency scaling diminishes, and the new misfit slowly transforms to an $\ell_{2}$-norm misfit.

In the second scenario, we test the robustness of the approach to coloured noise. For both inversions with noisy observed data in the absence of low frequencies below $3$ and $5$ Hz, $\ell_{2}$-norm did not successfully recover the Marmousi model due to cycle skipping and noise in the observed data as illustrated in Figures \ref{Inversion}(e)(i). Meanwhile, the proposed method demonstrates reasonable robustness towards noise by recovering the fine-scale parts of the true model, as illustrated in Figures \ref{Inversion}(f)(j). However, we observe some minor artefacts with additional "grainy" updates on the inverted model. Such defects can be attributed to the same properties that made the new misfit function work. Since the proposed misfit function boosts low frequencies in the waveform during its computation, it will eventually boost the coloured noise, which lies in a similar frequency spectrum, resulting in high energy noise in the updates. However, this might be alleviated by applying smoothing to the gradients in the updates. Apart from this, we also observed reduced accuracy along the boundary and the bottom region of the Marmousi model. This reduced accuracy is expected as the data coverage does not allow us to illuminate the deeper parts of the model, and the recorded signals are contaminated with noise (e.g., ambient noise) as later recording time. These challenges can be overcome with a far-offset acquisition network and high-quality data preprocessing before the inversion.

Despite the challenges, considering that the initial velocity is far away from the true model, the presence of noise, and the absence of low frequencies in the observed data, the proposed method has shown its potential in mitigating the cycle skipping issue and recovering the target model well.

\section{Conclusion}
We introduced \emph{$\mathbf{\omega}$-\textbf{FWI}}, a novel inversion methodology based on the \emph{Fourier-based metric} and demonstrated its robustness in full-waveform inversion. The proposed metric allows us to update the smoother components of the model (fit the low frequencies) naturally while admitting high-resolution updates as we get closer to the true solution. We tested the approach on the Marmousi data, absent low frequencies, and it produced a high-resolution result. Further application on a noisy version of the Marmousi data demonstrated its reasonable robustness.
\section*{Acknowledgments}
The first author would like to acknowledge Matteo Ravasi and Abdullah Alali from KAUST, and David Vargas from Utrecht University for the fruitful discussion and advice on waveform modelling and inversion. This publication is based on work supported by King Abdullah University of Science and Technology (KAUST).

\printbibliography

\end{document}